\documentclass[aps,twocolumn,prl,floatfix,showpacs,superscriptaddress]{revtex4}

\usepackage{amssymb,amsmath}
\usepackage{graphics,epsfig}

\begin{document}

\title{Dynamic Model of Fiber Bundles}

\author{M.Y. Choi}
\affiliation{Department of Physics, Seoul National University,
Seoul 151-747, Korea}
\affiliation{Korea Institute for Advanced Study,
Seoul 130-012, Korea}

\author{J. Choi}
\affiliation{Department of Physics, Keimyung University, Taegu
704-701, Korea}

\author{B.-G. Yoon}
\affiliation{Department of Physics, University of Ulsan, Ulsan
680-749, Korea}


\begin{abstract}
A realistic continuous-time dynamics for fiber bundles is introduced and
studied both analytically and numerically.  The equation of motion reproduces 
known stationary-state results in the deterministic limit
while the system under non-vanishing stress always breaks down 
in the presence of noise. 
Revealed in particular is the characteristic time evolution that
the system tends to resist the stress for considerable time, 
followed by sudden complete rupture.
The critical stress beyond which the complete rupture emerges is also obtained.
\end{abstract}

%

\maketitle


During past decades fiber bundle models have received considerable
attention and been studied extensively~\cite{sil}.  Originally
introduced to explain ruptures in heterogeneous material under
tension~\cite{hd}, fiber bundle models have been applied to
cracks and fractures, earthquakes, and other related
breakdown phenomena~\cite{asl,sor1,zrsv}.  
The main issue in the studies of fiber bundles is the life time of 
a bundle under given applied stress and accordingly, 
how the system fails as the applied stress is gradually increased~\cite{lt}.  
Recent works have been extended to the study of thermodynamic transitions~\cite{pt},
damping effects of viscous fibers~\cite{khhp}, 
and thermally activated failures~\cite{therm}. 
%
The breaking dynamics in the fiber bundle model is essentially determined 
by the condition whether the threshold of a particular fiber is exceeded by 
the applied stress and the equation of motion is expressed as recursion
relations.  It is thus deterministic, even though the
thresholds of the fibers are drawn from some probability
distributions which may include various effects~\cite{bpc}.  
The recursion equations can be solved 
by means of the graphical method and have been shown to 
be useful in classifying all possible threshold distributions.
Here time is viewed as a discrete variable and it is assumed that at each 
time step all the fibers meeting the breaking condition break immediately.  
The concern is then the number of intact fibers in the stationary state 
after evolving in discrete time.  Unlike the stationary state, however, 
this deterministic recursive dynamics, based on synchronous updating of 
the fibers at each {\em discrete} time step, does not provide a realistic
description of the actual dynamics.  In real systems it is obvious that
time is {\em continuous} and there usually exists retardation.  
Namely, it may take time for a fiber to respond to a given
stress and for the stress to get redistributed among fibers. 

This work is the first attempt toward realistic dynamics of fiber bundles, 
incorporating those features together with the probabilistic generalization. 
This kind of continuous-time dynamics in general yields the equations of motion
in the form of delay-differential equations and was successfully applied to 
neural network problems~\cite{myc}. 
From this formulation, we obtain the equation of motion for the average 
number of intact fibers, and find that the stationary solutions for 
simple threshold distributions are the same as those found previously in 
the recursion-equation approach. 
For realistic distributions,
the equation is solved numerically to give the time evolution. 
It is revealed that after initial rupture the average number of intact 
fibers tends to remain more or less constant, exhibiting plateau-like
behavior as a function of time.  In other words, even in case that the stress 
is strong enough to break all the fibers eventually, the system resists the stress for 
considerable time before complete rupture.
The critical stress beyond which the complete rupture emerges is also obtained.

We consider a bundle of $N$ fibers pulled by an external force $F$. 
In the global load-sharing limit, all fibers share the force uniformly so that 
which provides stress to each fiber. 
Each fiber has its own threshold and resists the stress lower than the threshold,
thus remaining intact. 
If the stress exceeds the threshold, however, the fiber becomes broken, 
leaving the applied stress redistributed among the neighboring intact fibers. 
For convenient representation, 
we assign a ``spin'' variable to each fiber in such a way that 
$s_i=+1\, (-1)$ for the $i$th fiber broken (intact). 
The state of the bundle is described by the configuration of all the
fibers, i.e., $\mathbf{s} \equiv (s_1, s_2,..., s_N)$.  
The total number of the intact fibers is related with the average spin
$\bar s \equiv N^{-1} \sum_j s_j$ via
\begin{equation}\label{intact}
N_{-}=\sum_{j=1}^N \frac{1-s_j}{2}=\frac{N}{2} (1- \bar s),
\end{equation}
and we are interested in how $N_-$ evolves in time as well as its 
stationary value.

The total stress on the $i$th fiber can then be written in the form
\begin{equation} \label{f}
 \eta_i = f + \sum_j V_{ij}\,\frac{1+s_j }{2},
\end{equation}
where $f$ is the stress directly due to the external force 
and $V_{ij}$ represents the stress transferred from the $j$th fiber 
(in case that it is broken). 
The breaking of the $i$th fiber with threshold $h_i$ is determined according to: 
\begin{eqnarray}
& \eta_i < h_i ~\Rightarrow~ s_i=-1 \nonumber\\&
\ \eta_i > h_i ~\Rightarrow~ s_i=+1 ,\label{dyn}
\end{eqnarray}
which, through the use of Eq. (\ref{intact}), can be simplified as
\begin{equation}\label{zerot}
   s_i E_i > 0
\end{equation}
with the local field $E_i \equiv (\eta_i - h_i)(1- \bar s)/2$. 
This determines the stationary configuration at zero ``temperature'' 
(i.e., in the deterministic limit).  

To describe the time evolution toward the stationary state described by 
Eq. (\ref{zerot}), we also take into consideration the uncertainty (``noise'')
present in real situations, which may arise from impurities and other 
environmental influences. 
We thus begin with the conditional probability 
that the $i$th fiber breaks at time $t{+}\delta t$, given that it is intact 
at time $t$:
\begin{equation}\label{con1}
    p(s_i{=}+1, t{+}\delta t | s_i{=}-1, t; \mathbf{s}', t{-}t_d)
        =\frac{\delta t}{2t_r}[1+\tanh \beta E'_i],
\end{equation}
where $\mathbf{s}' \equiv (s'_1, s'_2, \ldots, s'_N )$ represents the configuration 
of the system at time $t{-}t_d$ and $E'_i \equiv (\eta'_i - h_i)(1- \bar s')/2$ is
the local field at time $t{-}t_d$.  Note the two time scales $t_d$ and $t_r$ here: 
$t_d$ denotes the time delay during which the stress is redistributed among fibers 
while the refractory period $t_r$ sets the relaxation time of the system.
The temperature $T\equiv \beta^{-1}$ measures the width of the threshold 
region of the fibers or the noise level~\cite{com}: In the deterministic limit $(T =0)$, 
which is our main concern, 
the factor $(1+\tanh \beta E'_i)/2$ in Eq. (\ref{con1}) reduces to the step function 
$\theta (E'_i)$, yielding the stationary-state condition given by Eq. (\ref{zerot}).
The conditional probability that the $i$th fiber is
reconnected given that it is broken at time $t$ is obviously zero~\cite{com1}
\begin{equation}\label{con2}
    p(s_i{=}-1, t{+}\delta t | s_i{=}+1, t; \mathbf{s}', t{-}t_d)
      = 0 .
\end{equation}
Equations (\ref{con1}) and (\ref{con2}) can be combined to give a general expression for
the conditional probability $p(s'_i, t{+}\delta t | s_i, t; \mathbf{s}', t{-}t_d)$, 
which, in the limit $\delta t \rightarrow\infty $, can be expressed in terms of 
the transition rate:
\begin{eqnarray} \label{cong}
 & & p(s'_i, t{+}\delta t | s_i, t; \mathbf{s}', t{-}t_d) \nonumber \\
 & & ~~~~ = \left\{\begin{array}{ll}
     w_i(s_i; \mathbf{s}', t{-}t_d )\delta t &~\mbox{for}~ s'_i =-s_i \\
     1-w_i(s_i; \mathbf{s}', t{-}t_d )\delta t &~\mbox{for}~ s'_i=s_i,
    \end{array} \right .
\end{eqnarray}
where the transition rate is given by
\begin{equation}\label{tran}
    w_i(s_i; \mathbf{s}', t{-}t_d )=\frac{1}{2t_r}
       \left[\frac{1-s_i}{2}+\frac{1-s_i}{2}\tanh \beta E'_i \right].
\end{equation}

The behavior of the fiber bundle is then governed by the master equation, 
which describes the evolution of the joint probability
$P(\mathbf{s},t; \mathbf{s}', t{-}t_d)$ that the system is in 
state $\mathbf{s}'$ at time $t{-}t_d$ and in state $\mathbf{s}$
at time $t$:
\begin{eqnarray} \label{nme}
& & P(\mathbf{s}, t{+}\delta t; \mathbf{s}', t{-}t_d)-
    P(\mathbf{s}, t; \mathbf{s}', t{-}t_d) \nonumber \\
& &~~ = -\sum_{\mathbf{s}''} 
     [p(\mathbf{s}'', t{+}\delta t| \mathbf{s}, t; \mathbf{s}', t{-}t_d )
      P(\mathbf{s}, t; \mathbf{s}', t{-}t_d ) \nonumber \\
& &~~~~~~  -p(\mathbf{s}, t{+}\delta t| \mathbf{s}'', t; \mathbf{s}', t{-}t_d)
      P(\mathbf{s}'', t; \mathbf{s}', t{-}t_d) ] .
\end{eqnarray}
Thus we obtain the equation of motion in the form of a non-Markov master equation. 
Here the conditional probability for the whole system is given by the product of that 
for each fiber
$$
p(\mathbf{s}'', t{+}\delta t| \mathbf{s}, t; \mathbf{s}', t{-}t_d ) =
\prod_{i=1}^{N} p(s''_i, t{+}\delta t | s_i, t; \mathbf{s}', t{-}t_d).
$$
In the limit $\delta t \rightarrow \infty $, Eq. (\ref{nme}) takes the
differential form:
\begin{eqnarray}\label{mast}
\frac{d}{dt} P(\mathbf{s},t; \mathbf{s}', t{-}1) 
 &=& -\sum_i[w_i(s_i; \mathbf{s}') P(\mathbf{s}, t; \mathbf{s}', t{-}1) \nonumber \\
  &-& w_i(-s_i; \mathbf{s}') P(F_i\mathbf{s}, t; \mathbf{s}', t{-}1)],
\end{eqnarray}
where time $t$ has been rescaled in units of the delay time $t_d$,
the transition rate is given by 
$w_i(s_i; \mathbf{s}')\equiv t_d w_i(s_i; \mathbf{s}', t{-}t_d)$
with $w_i(s_i; \mathbf{s}', t{-}t_d)$ defined in Eq. (\ref{tran}),
and $F_i\mathbf{s}\equiv (s_1, s_2,..., -s_i, s_{i+1},..., s_N)$.
Then equations describing the time evolution of relevant physical
quantities in general assume the form of differential-difference
equations due to the retardation in the stress redistribution. In
particular, the average spin for the $k$th fiber,
$ m_k(t)\equiv
    \langle s_k \rangle \equiv\sum_{\mathbf{s},\mathbf{s}'}s_k
    P(\mathbf{s},t;\mathbf{s}', t{-}1)$
can be obtained from Eq. (\ref{mast}), by multiplying $s_k$ and
summing over all configurations:
\begin{eqnarray}
 \frac{d}{dt}m_k (t) 
  &=& - \sum_{\mathbf{s}, \mathbf{s}'}[s_k w_k(s_k; \mathbf{s}')
       P(\mathbf{s}, t; \mathbf{s}', t{-}1) \nonumber \\
  & &~- s_k w_k(-s_k; \mathbf{s}') P(F_k \mathbf{s}, t; \mathbf{s}', t{-}1)]\nonumber \\
  &=& -2\langle s_k w_k(s_k; \mathbf{s}') \rangle ,\label{mag}
\end{eqnarray}
where $\langle \;\rangle$ denotes the average over
$P(\mathbf{s}, t; \mathbf{s}', t{-}1)$. 
Evaluation of the average $\langle s_k w_k(s_k; \mathbf{s}') \rangle$, 
with $s_k^2=1$ noted, leads to
\begin{equation}
\tau \frac{d}{dt}m_k =
\frac{\displaystyle {1-m_k}}{\displaystyle 2}+
    \left\langle\frac{\displaystyle {1-s_k}}{\displaystyle 2}\tanh \beta E'_k \right\rangle
  \label{eqn}
\end{equation}
where $\tau\equiv t_r/t_d$ gives the relaxation time (in units of $t_d$). 

To proceed further, we need to specify the explicit form of $V_{ij}$. 
In the simplest case of global load sharing, which has been mostly studied
due to the analytical tractability, we have $V_{ij} = \eta_j /N_-$ as well as
$f=F/N$. 
Equation (\ref{f}) then leads to $(1-\bar s)\eta_i = 2f$ and accordingly,
the local field $E_i = f - (h_i /2)(1-\bar s)$. 
The infinite-range nature of such global load sharing allows one
to replace $E'_k$ by its average
$\langle E'_k\rangle = f - (h_k /2)[1-\bar m(t{-}1)]$,
where it has been noted that $s'$ is the configuration at time $t{-}1$,
i.e., $\langle\bar s'\rangle =N^{-1}\sum_{j}\langle\bar
    s'_j \rangle =N^{-1}\sum_{j}m_j (t{-}1) \equiv \bar m(t{-}1)$.

For convenience, we now rewrite Eq. (\ref{eqn}) in terms of the
average number of intact fibers at time $t$. 
Defining $x_k \equiv (1-m_k)/2 $ and 
$\bar x\equiv N^{-1}\sum_{k}x_k = (1-\bar m)/2$,
we have, from Eq. (\ref{intact}), 
$
\langle N_{-}\rangle = 
N \bar x 
$
and thus obtain from Eq. (\ref{eqn}) the equation of motion for the average fraction 
of intact fibers:
\begin{equation}\label{xk}
    \tau \frac{d}{dt}x_k (t)=-x_k (t)-x_k(t)\tanh \beta [f {-}h_k \bar x(t{-}1)]
\end{equation}
or, upon the sample average, 
\begin{equation}\label{sxav}
\tau \frac{d}{dt} \bar x(t)=-\bar x(t)- \langle\!\langle
x(t)\tanh \beta [f {-}h_k \bar x(t{-}1)]\rangle\!\rangle,
\end{equation}
where $\langle\!\langle \;\rangle\!\rangle$ stands for the average
with respect to the distribution of $\{h_k\}$ and self-averaging
has been assumed.

We first examine the stationary solutions of Eq. (\ref{xk})
\begin{equation}\label{sta}
    x_k +x_k \tanh \beta [f {-}h_k \bar x)]=0,
\end{equation}
which possesses two possible solutions: 
$x_k =0$ or $\tanh \beta (f{-}h_k \bar x)=-1$. 
The latter is possible only if $f < h_k \bar x$ at $T=0$. 
In this case, the stress $f$ is not large enough to break the fiber, 
so that we have $x_k=1$.  
When $f > h_k \bar x $, on the other hand, $x_k=0$ is the only solution. 
It is thus concluded that 
\begin{equation}\label{sol}
    x_k =\theta (h_k \bar x {-}f)
\end{equation}
in the deterministic limit ($T=0$). 
The average fraction of intact fibers,
given by the sample average, i.e., the average over the distribution of the 
threshold, then becomes
\begin{equation}\label{solbar}
    \bar x=\int dh g(h)x =\int dh \theta (h \bar x {-}f )g(h),
\end{equation}
where $g(h)$ is the distribution function of the threshold $\{h_k\}$. 
This is the self-consistent equation, the solution of which gives the
stationary fraction of the surviving fibers. 
Note also that the stationarity condition 
$\bar x+ \langle\!\langle x \tanh \beta [f {-}h_k \bar x]\rangle\!\rangle=0$
or
\begin{equation}\label{cond}
    \int dh g(h) \theta (h \bar x {-}f )[1+\tanh \beta (f {-}h\bar x)]=0
\end{equation}
is automatically satisfied since $\tanh \beta(f{-}h \bar x)
=\theta(f{-}h \bar x)- \theta (h \bar x {-}f)$. 
At $T\neq 0$, we have $x_k =0$, giving $\bar x=0$ as the
only possible solution.  

We now briefly discuss how to solve the Eq. (\ref{solbar}). For a
continuous distribution of the threshold, we formally solve Eq.
(\ref{solbar}) by simply performing the integration
\begin{equation}\label{solu}
    \bar x= 1-G(f /\bar x),
\end{equation}
where $G(h)=\int_0^h dh' g(h')$ is the cumulative distribution of
the threshold.  This equation corresponds to the fixed-point 
equation for the deterministic dynamics, except that in our
case $\bar x$ is a double averaged quantity. 
This equation may be solved, e.g., graphically for given $g(h)$ \cite{sil}, 
and yields all known results, which we do not reproduce here. 
For example, the simple bimodal distribution 
$g(h)=\rho \delta (h{-}f_1) +(1-\rho)\delta (h{-}f_2)$ with $0<f_2<f_1$ 
and $0\leq \rho\leq 1$ leads to
\begin{equation}\label{mod}
   \bar x=\int dh \theta (h\bar x {-}f )[\rho\delta (h{-}f_1) +(1-\rho)\delta (h{-}f_2)]
\end{equation}
whose solution is given by
\begin{equation*} 
\bar x
  = \left\{\begin{array}{ll}
     1 &~\mbox{for} ~ f < f_2 \bar x \\
     \rho &~\mbox{for} ~ f_2 \bar x < f < f_1 \bar x \\
     0 &~\mbox{for} ~ f_1 \bar x < f .
    \end{array} \right.
\end{equation*}
As $f$ is increased from zero, the first rupture occurs at
$f =f_2$, yielding $\bar x=\rho$, and the second one occurs at
$f =\rho f_1$ if $f_2 < \rho f_1$.  For $f_2 > \rho f_1$, however,
there appears only one rupture at $f =f_2$. 
In these two cases the critical stress $f_c$, beyond which the system breaks completely, 
is thus given by $\rho f_1$ and $f_2$, respectively.

\begin{figure}
\centerline{\epsfig{file=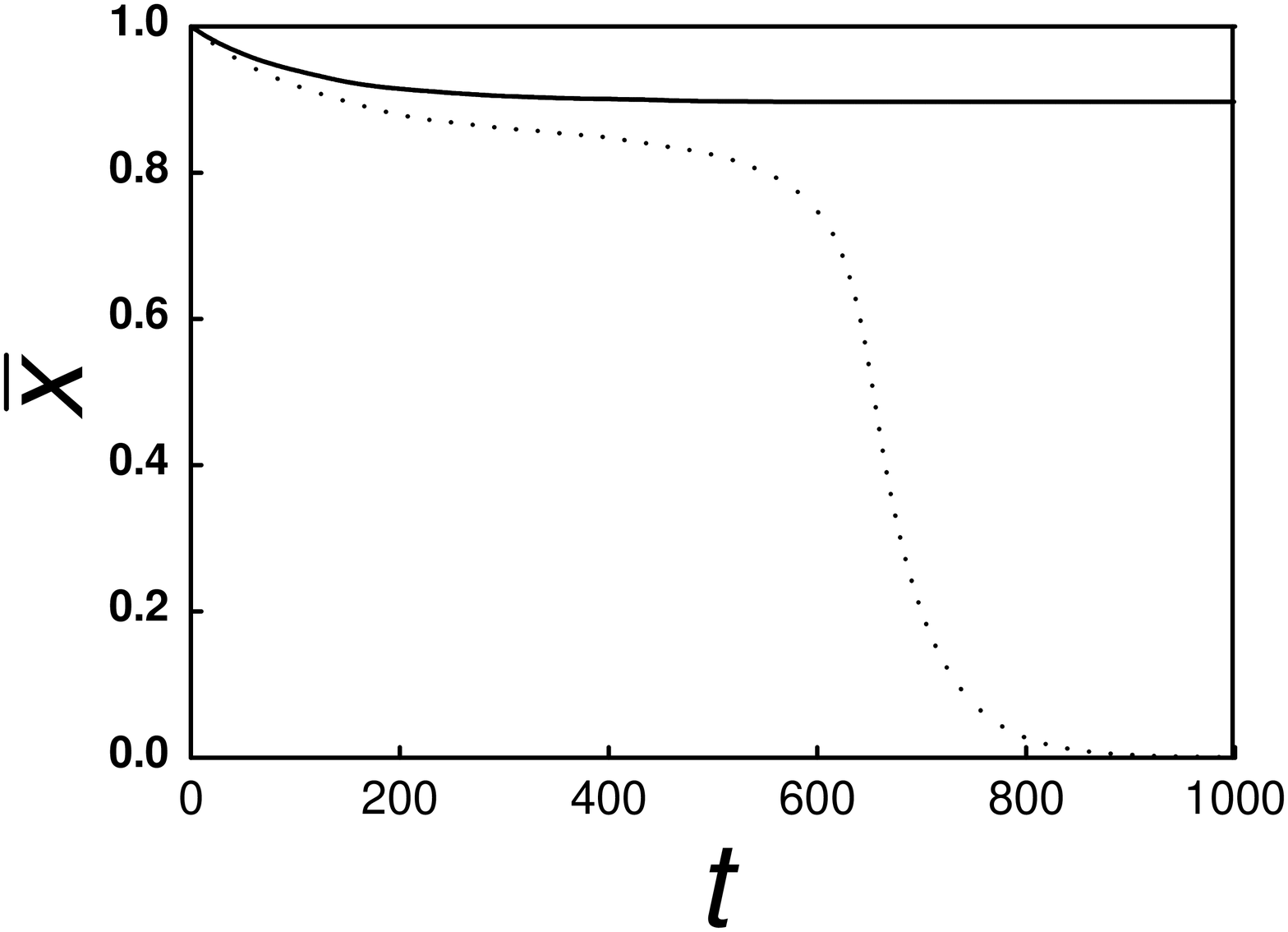,width=7cm}}
\caption{Behavior of the average fraction $\bar x$ of intact fibers with time $t$ 
(in units of the delay time) for the Gaussian distribution of $\bar h =1$ and 
$\sigma =0.2$. Solid and dotted lines correspond to $f=0.66$ and $0.68$, respectively.
}
\label{fig:zero}
\end{figure}
Next the time evolution of the system is explored via direct numerical integration of
the equation of motion (\ref{xk}).  
We consider realistic distributions of the threshold including
the Gaussian distribution as well as the uniform one.
Specifically, we set $\tau=100$ and use the time step $\Delta t =0.1$, mostly 
in a system of $N=10^3$ fibers.  These parameter values have been varied, only to 
give no appreciable difference.

\begin{figure}
  \centerline{\epsfig{file=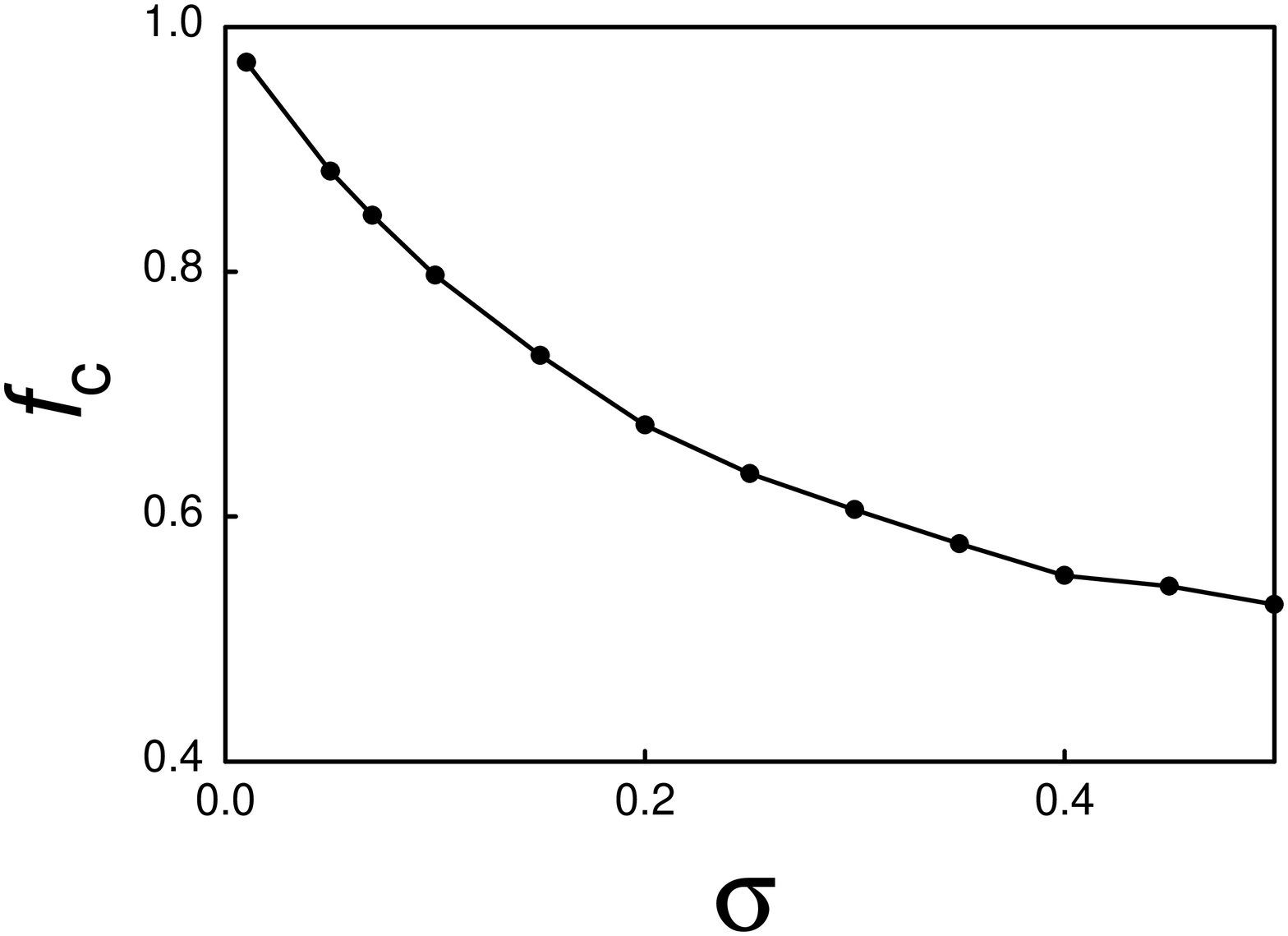,width=7cm}}
  \caption{Critical stress $f_c$ versus variance $\sigma$}
  \label{fig:fc}
\end{figure}
The typical behavior of the average fraction $\bar x$ of intact fibers is shown in
Fig. \ref{fig:zero} for the Gaussian distribution with unit mean $(\bar h=1)$
and variance $\sigma =0.2$.  Note that for $f = 0.66$ a large part of the fibers 
remain intact whereas all fibers become broken for $f =0.68$, indicating that
the critical stress $f_c$ lies in between.
Indeed this is the case in Fig. \ref{fig:fc}, which displays how the critical stress 
varies with the variance of the threshold distribution. 
It is of interest that after initial rupture $\bar x$ tends not to change much 
and exhibits a plateau as a function of time, 
which persists regardless of the details of the threshold distribution. 
This indicates that even for the stress 
strong enough to break all the fibers eventually, the system resists the stress for 
considerable time, followed by the complete rupture occurring suddenly. 
Note the sharp contrast with the results based on the conventional discrete-time 
recursive dynamics, where the strong stress brings about immediate failure \cite{bpc}. 

\begin{figure}
  \centerline{\epsfig{file=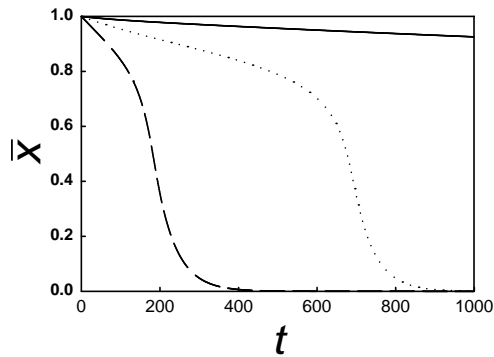,width=7cm}}
  \caption{Behavior of the average fraction of intact fibers with time at $T=0.1$
  for $f = 0.48$\,(solid line), $0.58$\,(dotted line), and $0.68$\,(dashed line).
  Other parameters are the same as those in Fig. \ref{fig:zero}.
   }
  \label{fig:nonzero}
\end{figure}

\begin{figure}
  \centerline{\epsfig{file=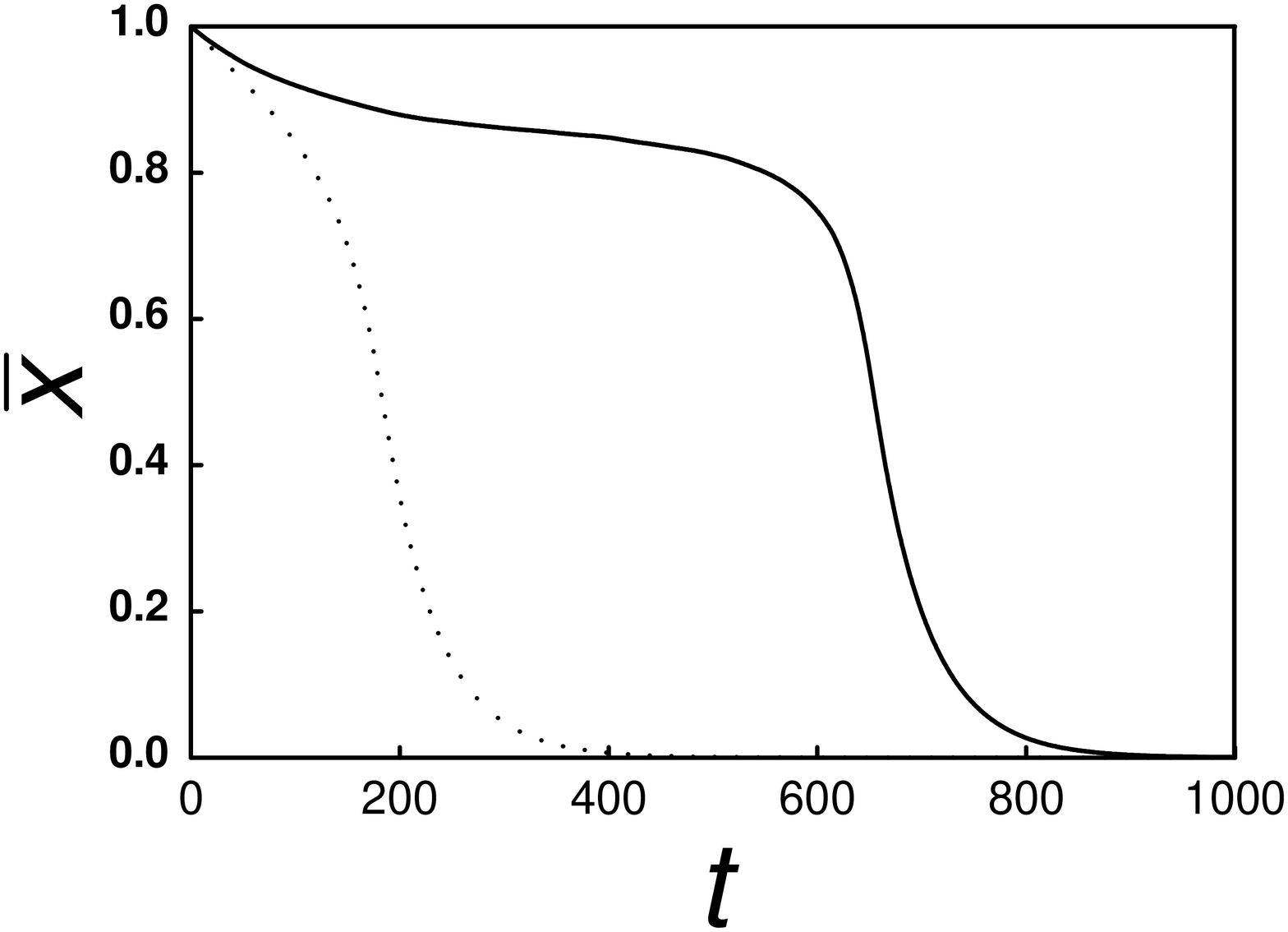,width=7cm}}
  \caption{Comparison of the behavior at $T=0$\,(solid line) and $0.1$\,(dotted line).
   The stress is $f =0.68$ and other parameters are the same as those in 
   Fig. \ref{fig:zero}. }
  \label{fig:compare}
\end{figure}
We also examine the effects of noise ($T\neq 0$), which disallows the 
nontrivial solution in the stationary state and leads to $f_c =0$. 
Nevertheless Fig. \ref{fig:nonzero} displays that the rupture time can be
very long unless $f$ is not large.  In a real system $T$ is nonzero
but usually very small, and the complete rupture practically does not occur
in the time scale of interest. 
For large $f$, the rupture is accelerated in the presence of noise,
as shown in Fig. \ref{fig:compare}.  

In summary, we have introduced realistic continuous-time dynamics for 
fiber bundles and investigated the behavior of the system under stress.
The dynamics has been generalized to include uncertainty
due to impurities and environmental influences.  In its presence 
the system has been found always to break eventually, reflecting the irreversible 
nature of breaking, whereas in its absence all stationary features found in the
deterministic recursion-equation approach have been reproduced. 
In particular disclosed is characteristic time evolution that the system tends to resist 
the stress for considerable time, followed by the complete rupture occurring suddenly. 
This has interesting implications to many systems in nature such as biological problems,
the investigation of which is left for further study. 

This work was supported in part by the Overhead Research Fund of SNU and in part by 
the Basic Research Program (Grant No. R01-2002-000-00285-0) of KOSEF.


\begin{thebibliography}{99}
\bibitem{com}
Here $T$ is not directly related with the thermodynamic temperature of the system,
considered in Refs. \onlinecite{therm} and \onlinecite{bpc}.
\bibitem{com1}
Then the problem corresponds to the Choi model of neural networks
with $t_r /t_0 \ \rightarrow 0$.  See  Ref. \onlinecite{myc}.
\end{thebibliography}
\end{document}